\documentclass[preprint,superscriptaddress,floatfix,sort&compress]{revtex4}

\usepackage[font=scriptsize]{caption}
\usepackage{color,graphicx}

\usepackage{ulem}

\begin{document}

\title{Simulating topological domains in human chromosomes with a fitting-free model}

\author{C.~A. Brackley}
\affiliation{SUPA, School of Physics \& Astronomy, University of Edinburgh, Peter Guthrie Tait Road, Edinburgh, EH9 3FD, UK}
\author{D. Michieletto}
\affiliation{SUPA, School of Physics \& Astronomy, University of Edinburgh, Peter Guthrie Tait Road, Edinburgh, EH9 3FD, UK}
\author{F. Mouvet}
\affiliation{SUPA, School of Physics \& Astronomy, University of Edinburgh, Peter Guthrie Tait Road, Edinburgh, EH9 3FD, UK}
\author{J. Johnson}
\affiliation{SUPA, School of Physics \& Astronomy, University of Edinburgh, Peter Guthrie Tait Road, Edinburgh, EH9 3FD, UK}
\author{S. Kelly}
\affiliation{Department of Plant Sciences, University of Oxford, South Parks Road, Oxford OX1 3RB, UK}
\author{P.~R. Cook}
\affiliation{Sir William Dunn School of Pathology, University of Oxford, South Parks Road, Oxford, OX1 3RE, UK}
\author{D. Marenduzzo}
\affiliation{SUPA, School of Physics \& Astronomy, University of Edinburgh, Peter Guthrie Tait Road, Edinburgh, EH9 3FD, UK}

\begin{abstract}
We discuss a polymer model for the 3D organization of human chromosomes. A chromosome is represented by a string of beads, with each bead being ``colored'' according to 1D bioinformatic data (e.g., chromatin state, histone modification, GC content). Individual spheres (representing bi- and multi-valent transcription factors) can bind reversibly and selectively to beads with the appropriate color. During molecular dynamics simulations, the factors bind, and the string spontaneously folds into loops, rosettes, and topologically-associating domains (TADs).
This organization occurs in the absence of any specified interactions between distant DNA segments, or between transcription factors.  A comparison with Hi-C data shows that simulations predict the location of most boundaries between TADs correctly. The model is ``fitting-free'' in the sense that it does not use Hi-C data as an input; consequently, one of its strengths is that it can -- in principle -- be used to predict the 3D organization of any region of interest, or whole chromosome, in a given organism, or cell line, in the absence of existing Hi-C data. We discuss how this simple model might be refined to include more transcription factors and binding sites, and to correctly predict contacts between convergent CTCF binding sites.
\end{abstract}

\maketitle

\section*{Hi-C: contact maps, domains and loops}

The conformations adopted by human chromosomes in 3D nuclear space are key contributors to gene activity in health and disease~\cite{Cavalli2013}, and understanding the principles driving genome folding is one primary goal of biophysicists studying DNA. An important recent experimental breakthrough has been the development of chromosome conformation capture (3C), and its high-throughput derivative -- ``Hi-C'' -- which allows contacts between different chromatin segments to be mapped genome-wide~\cite{Lieberman-Aiden2009,Dixon2012,Rao2014}. 

Contact maps obtained using Hi-C reflect some underlying chromosomal organization. For example, each chromosome folds into distinct ``topologically-associating domains'' (TADs) during interphase (but not during mitosis when transcription ceases~\cite{Naumova2013}). Domain size is variable, with higher-resolution studies typically uncovering smaller TADs in the range between 0.1-2 Mbp~\cite{Dixon2012,Rao2014}. TADs are largely specified by the local chromatin environment, as the same 20-Mbp region in a chromosomal fragment or an intact chromosome yield similar contact maps~\cite{Rao2014}. This organization into TADs is conserved, as they are found in budding yeast~\cite{Hsieh2015} and {\it Caulobacter crescentus}, where they are called ``chromosomal interaction domains'' or CIDs~\cite{Le2013}. CIDs are also separated by strong promoters, and they are eliminated by inhibiting transcription.

Bioinformatic analysis suggests that eukaryotic TADs tend to be epigenetically determined; active and inactive regions typically form separate domains~\cite{Lieberman-Aiden2009,Dixon2012,Rao2014,Sexton2012}, with CTCF (the CCCTC-binding transcription factor) and active transcription units (or binding sites for RNA polymerase II) being enriched at inter-domain ``boundaries''~\cite{Rao2014,Dixon2012}. These analyses also uncover chromosome loops apparently stabilized by transcription factors bound to promoters and enhancers~\cite{Rao2014,Mifsud2015,Simonis2006,Li2012,Jin2013,Zhang2013,Heidari2014}, or CTCF bound to its convergent cognate sites (presumably the latter loops are tethered by associated cohesin complexes acting as a molecular ``slip-link'', or ``hand-cuff'')~\cite{Dixon2012,Rao2014}. Remarkably, many fewer loops are associated with divergent or parallel CTCF binding sites~\cite{Rao2014}.
 
While Hi-C data is normally obtained using cell populations of milions of cells, single-cell Hi-C experiments show that no two cells in the same population share exactly the same contacts; nevertheless, the organization is non-random as certain contacts are seen more often than others~\cite{Nagano2013}.

These observations point to central roles for transcription orchestrating the 3D organization of chromosomes, with transcription factors providing molecular ties which stabilize the structure both locally and globally. The results also suggest that CTCF and cohesin are important organizers, with the latter providing an example of a molecular slip-link. 
Here we discuss results obtained using a simple biophysical model, which is based on the binding of two types of transcription factors to cognate sites on DNA. As we will see, molecular dynamics simulations using this model yield contact maps remarkably similar to those obtained from Hi-C.  We further discuss how this model can be extended to incorporate more transcription factors, and molecular slip-links like cohesin.

\section*{A toy model, and some basic principles}

We first introduce a toy model which is schematically described in Figure 1A: a chromatin fiber (represented by a flexible bead-and-spring chain) interacts non-specifically with bi- or multi-valent spheres (this toy model is analogous to the ``strings-and-binders'' model of~\cite{Nicodemi2009BJ,Barbieri2012}). The red spheres in Figure 1A represent transcription factors or complexes that can bind to two or more sites on the fiber; consequently they can form ``molecular bridges'' that stabilize loops. These factors stick to the chromatin fiber via a generic attractive interaction. If the interaction strength is large enough to allow multivalent binding, then the bound proteins spontaneously cluster, a phenomenon first observed and discussed in~\cite{Brackley2013}. This clustering is accompanied by the formation of chromatin ``domains'', in which intra-domain contacts are enriched over inter-domain ones. The (generic) principle underlying clustering -- which occurs in the complete absence of any specified DNA-DNA or protein-protein interaction -- has been called the ``bridging-induced attraction''-- as it does not occur with univalent proteins that cannot stabilize loops~\cite{Brackley2013,Brackley2016NAR,LeTreut2016,Michieletto2016}. 

The basic mechanism underlying this attraction is a simple thermodynamic positive feedback loop (Fig. 1B). First, proteins bind to chromatin, and -- as they are at least bivalent -- they can form a molecular bridge between two different DNA segments. This bridging brings distant parts of the chromosome together to increase the local chromatin concentration; this makes it more likely that additional proteins in the soluble pool will bind as they diffuse by. And once they have bound, these proteins will form additional molecular bridges which increase the chromatin concentration further. As this cycle repeats, protein clusters form, and these nucleate TAD-like structures. [We assume that the protein concentration is sufficiently low that proteins cannot completely cover the fiber even when all bind. If, instead, the protein concentration is very large, then bridging induces macroscopic collapse of the whole fiber~\cite{Nicodemi2009BJ,Barbieri2012,Johnson2015}.]

In this simple case in which the transcription factors only bind non-specifically, the bridging-induced attraction yields clusters that continue to grow in size, ultimately giving one single cluster in steady state~\cite{Johnson2015}. However, most transcription factors also bind specifically, as well as non-specifically. A simple modification of the toy model includes a stronger specific binding (of, e.g., red proteins to pink chromatin beads in Fig. 1C). Clusters still form via the bridging-induced attraction (Fig. 1C), but now they no longer grow indefinitely; instead, they reach a self-limiting size. This is because clustering of specifically-bound beads creates rosettes, or other structures with many chromatin loops, and bringing these together is entropically costly. Crucially, the entropic cost rises super-linearly with loop number, and this arrests cluster growth~\cite{Brackley2013,Brackley2016NAR,Marenduzzo2009}. 

Another simple consequence of this generic organizing principle is that multivalent binding naturally creates ``specialized'' clusters. Imagine that two types of transcription factor (i.e., ``red'' and ``green'') bind specifically to different beads on the fiber (i.e., pink and light green; Fig. 1C). Then, the bridging-induced attraction works for the red and green factors separately. For instance, red factors increase the local concentration of pink chromatin binding sites, this recruits more red proteins, etc. Consequently, the clusters that emerge tend to contain either red factors plus pink beads or green factors plus light-green beads. If red and green proteins represent complexes containing RNA polymerase II and III respectively, this naturally explains why distinct foci/``factories'' are seen in human cells that contain one or other enzyme, but not both~\cite{Xu2008}. As discussed in the next Section, a similar mechanism probably underlies the organization of the ``A/B'' compartments uncovered in Hi-C experiments~\cite{Lieberman-Aiden2009}.

\section*{A minimal, fitting-free, polymer model for chromosome folding}

The toy model of Figure 1 was extended in~\cite{Brackley2016NAR} to give a minimal fitting-free predictive model for genome organization. The model is fitting-free because it is based solely on 1D information on the protein binding (or epigenetic) landscape . Thus, unlike other commonly-used approaches, it does not rely on contact information as an input, so its predictive power is enhanced. In the version proposed in~\cite{Brackley2016NAR}, the whole of chromosome 19 in GM12878 cells was modeled (Fig.  2A). In this case, each chromatin bead contained 3 kbp, and factors were of two types -- ``active'' (modeling complexes of polymerases and transcription factors) or ``inactive'' (modeling heterochromatin-associated proteins like HP1$\alpha$, or even a simple linker histone like H1 -- as both proteins are known to bind the genome in multiple places~\cite{Kilic2015,Mack2015}). Beads in the chromatin fiber are ``colored'' according to bioinformatic data to specify whether they bind the active or inactive proteins. Thus, active beads were colored using the ``active'' Broad ChromHMM tracks~\cite{Ernst2011}~\footnote{The Broad ChromHMM track is available on the UCSC Genome Browser. To build it, several data sets for histone modification and protein binding have been analyzed using a hidden Markov model to classify chromatin regions as being in one of several chromatin ``states''.} on the hg19 assembly (i.e., using states 1,4,5 in the HMM track that signify an ``Active Promoter'' or ``Strong Enhancer'' to specify strong binding, and states 9 and 10 that signify ``Transcriptional Transition'' or ``Transcriptional Elongation'' to specify weak binding). Inactive beads were colored using either the appropriate HMM tracks or GC content -- the latter is illustrated here as a low GC content is such a good predictor of an inactive (heterochromatic) nature.

Given the simplicity of this model, it is striking to see how well it allows correct prediction of the positions of TADs and their boundaries (Fig. 2C,D). For example, 85\% boundaries are correctly identified to within 100 kbp; some inter-domain interactions are even correctly captured (see the off-diagonal blocks in the contact maps). While this agreement can certainly be improved by adding biological detail, we stress that it is especially remarkable as it appears in a fitting-free minimal model (the only relevant parameters are interaction strengths and cut-offs, but little difference is found if these are set to ensure multivalent binding). The model can be applied, in principle, to any chromosome for which appropriate bioinformatic data is available (e.g., Broad ChromHMM track or histone modifications~\cite{Brackley2016GB}); consequently, it can be used genome-wide in different cell lines and organisms. It can also be used to predict the contact map of any region of interest, and -- of course -- it can be applied at a higher resolution~\cite{Brackley2016GB}.

As in the toy model, active and inactive factors (and their cognate biding sites) cluster separately, and the model naturally yields the A (active) and B (inactive) compartments seen in Hi-C contact maps. Moreover, the proteins cluster to give structures reminiscent of both nuclear ``bodies'' (e.g., Cajal, polycomb and promyelocytic leukemia bodies), and factories containing RNA polymerases II and III -- all structures rich in distinct proteins binding to different DNA sequences~\cite{Sleeman2014,Pombo1999,Cook1999,Papantonis2013}. The number of protein clusters is significantly smaller than that of chromatin domains: therefore our model predicts that a number of TADs will come together into a single protein cluster (say, a transcription factory), but different TADs might interact in different cells. 

As these simulations reproduce the overall Hi-C organization well, it is of interest to ask what is special about beads at, or close to, boundaries between TADs.  Figure 3 shows that the boundary beads {\it in silico} are depleted of inactive beads and enriched in active marks: this is consistent with bioinformatic analyses showing that boundaries are depleted in heterochromatic marks like HK39me3 and K3K27me3, and enriched in active ones like H3K4me3, as well as in transcription start sites and binding sites for RNA polymerase II~\cite{Dixon2012}. An intriguing additional signal is that beads enriched at boundaries in silico are often non-binding beads -- which naturally form boundaries as they possess few contacts; this is consistent with ~15\% Hi-C boundaries lacking any particular mark~\cite{Dixon2012}. Finally, we note that, by using toy models, Refs.~\cite{Brackley2016NAR,Benedetti2014,Hofmann2015} showed that (permanent) chromatin loops (e.g., maintained by CTCF) may also act as boundaries, whose strength varies according to the force field used. This finding may be the reason why active beads are enriched at boundaries (they often constitute the base of loops, although these are dynamic ones).

\section*{Beyond the minimal model: adding colors and slip-links}

The minimal model described this far generally yields contact maps like those obtained from Hi-C data~\cite{Brackley2016NAR}; however, exceptions do exist. In general, the percentage of TAD boundaries predicted accurately increases with transcriptional activity (the organization of chromosome 19 is predicted well, perhaps because it is the one containing the most active genes). In less-active regions, boundaries are sometimes predicted less accurately: e.g., Fig. 4A shows a region, in chromosome 14, where the minimal model fails at correctly predicting the location of some TADs (most of which are inactive). This raises the questions whether it is possible to improve the ``coloring'' of inactive beads, and/or add more colors. Capture Hi-C results provide a way of adding more colors. Thus, Mifsud et al.~\cite{Mifsud2015} distinguished contacts between promoters on the basis of their histone marks, and found there chromatin regions bearing the H3K9me3 or the H3K27me3 mark interacted with other regions with the same mark, whereas ``mixed'' contacts between K27 and K9 trimethylated regions were very rare. H3K9me3 binds HP1 to yield constitutive heterochromatin~\cite{Lehnertz2003}. H3K27me3 is a classic inactive mark associated with facultative heterochromain and binding of polycomb-group repressing complexes; it marks ``blue chromatin'' in Drosophila~\cite{Filion2010}. Therefore, we improved our model by stipulating that heterochromatic beads were classified according to histone modifications (instead of GC content), with two different colors for beads bearing the H3K9me3 or H3K27me3 mark~\footnote{In practice, we used a threshold in histone modification tracks to color beads, but the exact value of the threshold played a minor role in the results.}; we then also included in the simulations two proteins binding to these marks (modeling, e.g., PcG-protein complexes, such as PRC1 binding to H3K27me3 marks~\cite{Michieletto2016,Wani2016}). 

Figure 4B shows that, once the two different heterochromatin beads are distinguished, the simulation predicts TAD patterns more accurately. We stress that the refined model is still fitting free as it does not rely on Hi-C data for input, but only assumes knowledge of 1D protein binding landscape, or histone modification profiles. 

Another (fitting-free) model similar in spirit to the one presented here is the ``block-copolymer'' model used to study folding of Drosophila chromosomes~\cite{Jost2014}.  [For a non-fitting free version, see~\cite{Giorgetti2014,Tiana2016}.] In this case, chromatin beads interact directly, so bridging proteins are implied but not explicitly modeled. This approach is equivalent to the one used in Figures 2-4 if bridging proteins are abundant enough to saturate binding sites; however, the two models differ in the regime where only some binding sites are occupied. The model used in Figures 2-4 also naturally explains the formation of nuclear bodies, and so can be used to study their biogenesis and kinetics (this is not possible with the block-copolymer model where  bridging proteins are assumed to be uniformly distributed at all times).

A recent study by Chiariello et al.~\cite{Chiariello2016} offers another avenue to improve simulation accuracy by using some information from Hi-C experiments (but then the model is not fitting free). In practice this is done through an iterative procedure which finds the minimal arrangement of binding sites and colors which best explain the Hi-C contact map; for example, simulations involving 16 colors gave contact maps for the Sox6 locus that were indistinguishable from those obtained by Hi-C (correlation coefficient 95\%).

An important unaddressed aspect concerns loops (or ``loop domains'') stabilized by CTCF~\cite{Rao2014}. As discussed above, CTCF is more likely to bridge two cognate binding sites~\cite{Rao2014,CRISPCTCF} when sites are in a ``convergent'' orientation compared to a ``divergent'' one. Polymer models to explain this have been proposed~\cite{LibAid2015,Mirny2016}; they involve loop-extrusion factors and slip-links that are simultaneously bound (linked) to beads on two different chromosomal segments and which can slide (slip) along the segments (in practice, these factors/slip-links are cohesin and/or condensin). These models can account for the observed CTCF orientation bias, as they assume that the loop-extrusion factors can stably stick only to one side of CTCF (which is true of cohesin). However, these models also require some as-yet undiscovered motor protein with a processivity sufficient to generate loops of hundreds of kb. Moreover, CTCF and its convergent sites cannot be the sole organizer of boundaries, as knock-outs of CTCF have only minor effects on domain organization in mammals~\cite{Zuin2014,Hou2010,Seitan2013}, and bacteria possess domains but no equivalent of CTCF. Nevertheless, CTCF directionality and cohesins clearly play an important role in the formation and establishment of several eukaryotic loops, so it will be of interest to incorporate these components into our model.

\section*{Acknowledgements}
CAB, DMi and DMa acknowledge support from ERC CoG 648050 (THREEDCELLPHYSICS).

\newpage

\begin{figure}
\centerline{\includegraphics[width=9.cm]{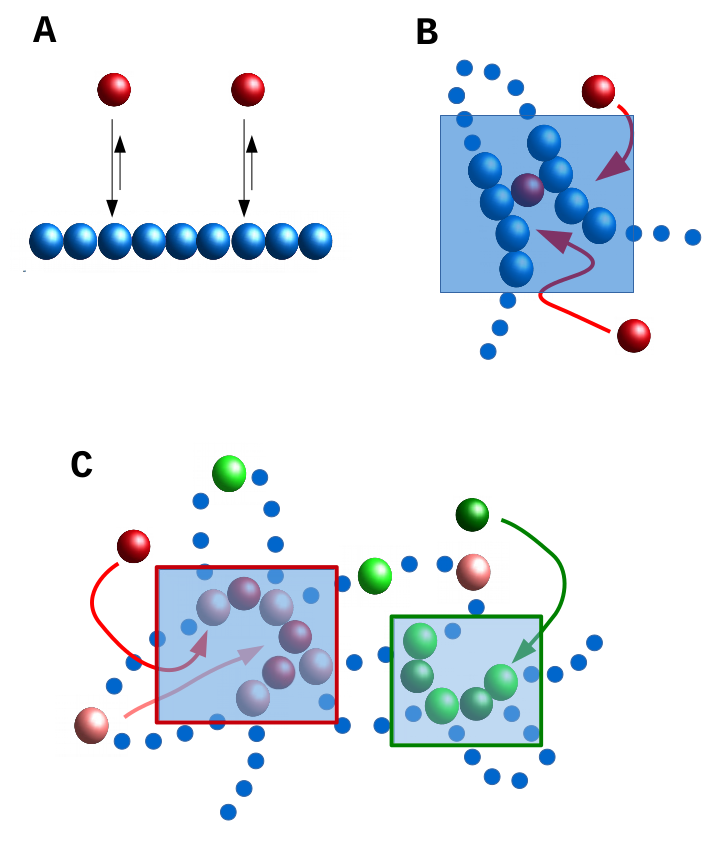}}
\caption{\label{Fig1}
Schematic representation of the toy model discussed in the text.
(A) A chromatin fiber is coarse-grained into a bead-and-spring polymer, where monomers are spherical (blue beads). Proteins (red beads) bind to the chromatin fiber non-specifically (arrows). (B) As proteins are multivalent, upon binding they can create molecular bridges: here the bound red protein contacts two blue chromatin beads, and this increases the local chromatin density (shaded area): therefore, other proteins in the soluble pool are more likely to bind chromatin in this area. This will, in turn, further increase chromatin density creating a (thermodynamic) positive feedback loop which eventually leads to the formation of protein clusters (concomitantly with TAD-like chromatin domains). (C) Schematic of a toy model with specific binding. Now red proteins bind specifically to pink chromatin beads, and green proteins to light-green chromatin beads. As proteins are multivalent, and because pink and light-green beads lie at different places along the fiber, a similar positive feedback as in (B) separately drives the increase of local concentration of pink and light-green chromatin beads (in the two shaded ares), which eventually leads to the formation of specialized clusters of red proteins and pink chromatin binding beads, and of green proteins and light-green binding beads.}
\end{figure}

\begin{figure}
\centerline{\includegraphics[width=9.cm]{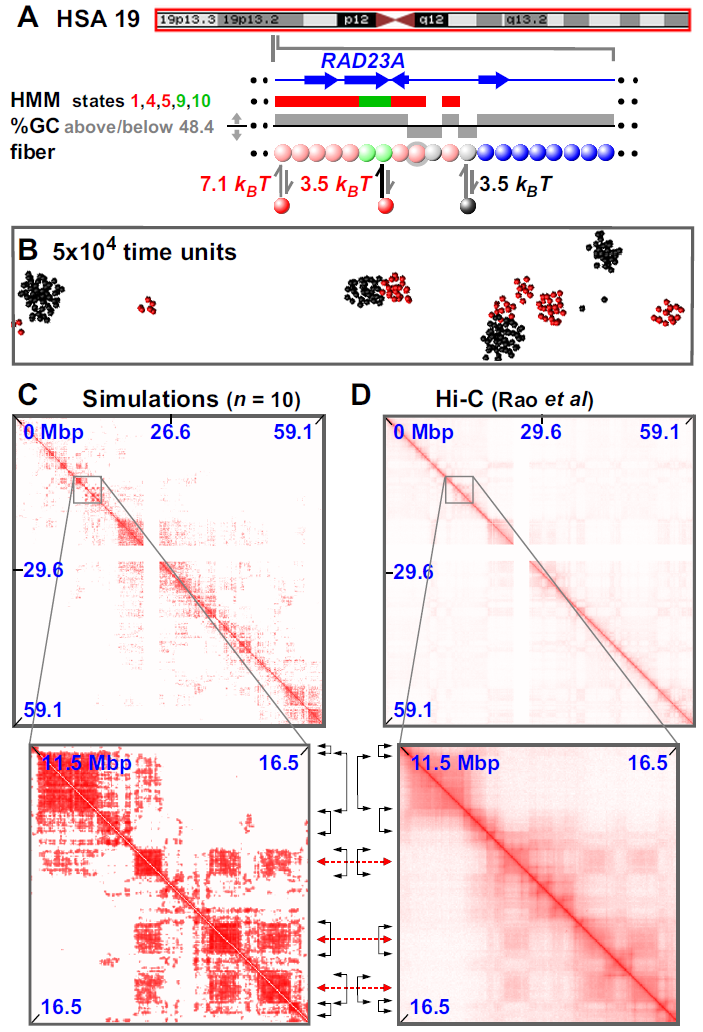}}
\caption{\label{Fig2}
Fitting-free simulations of chromosome 19 in GM12878 cells. 
(A) Overview. The ideogram (red box indicates the whole chromosome that was simulated) and Broad HMM track (colored regions reflect chromatin states) are from the UCSC browser; the zoom illustrates an arbitrary region, around {\it RAD23A}, to show the details of the ``coloring''. Beads (3 kbp) are colored according to HMM state and GC content: blue beads are non-binding; pink beads correspond to states 1,4,5 in the ChromHMM track; light-green to states 9,10. Grey beads correspond to beads which have $<$48.4\% GC. Pink and light-green beads bind (respectively, strongly and weakly) active factors (red in the figure); grey beads bind to inactive factors, linked to heterochromatization (black in the figure). Note that the coloring rule is such that beads can have multiple colors: for instance, in the zoom two pink beads are also grey (represented by grey halos), so that such beads can bind both red and black factors.
(B) Snapshot (without chromatin) of central region after 5$\times$10$^4$ units; most clusters contain factors (or proteins) of one color. In other words, active and inactive proteins cluster separately. As discussed in the text, the formation of specialised clusters may underlie both the formation of A/B compartments (when looking at the chromatin interactions) and that of some nuclear bodies (when looking at the protein cluster patterns).
(C,D) Comparison between contact maps from simulations and experiments (see Ref.~\cite{Brackley2016NAR} for more details). Between zooms, black double-headed arrows mark boundaries of prominent domains (on the diagonal), and red double-headed ones the centers of off-diagonal blocks making many inter-domain contacts. 
Reproduced from Ref.~\cite{Brackley2016NAR}, with permission.}
\end{figure}

\begin{figure}
\centerline{\includegraphics[width=10.cm]{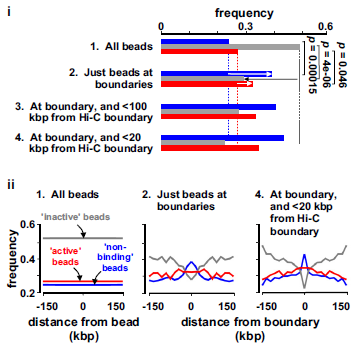}}
\caption{\label{Fig3}
Characterization of TAD boundaries found {\it in silico}.
These plots are obtained by analyzing the TAD boundaries found in simulations (through combination of an automated method and visual inspection~\cite{Brackley2016NAR}), and by computing the frequencies of non-binding (blue bars), inactive (grey bars) and active (red bars) beads in different sets.
 Set 1: all beads. Set 2: Beads lying within 100 kbp of a boundary. Sets 3 and 4: The sub-sets of set 2 that also lie within 100 and 20 kbp of a boundary identified in Hi-C data. 
(i) Beads at boundaries are rich in active and non-binding beads, and depleted of inactive beads (arrows; p values assessed assuming Poisson distributions). 
(ii) The frequencies of different beads (in sets 1, 2 and 4) in the 150 kbp on each side of either each bead in set 1, or of boundaries in sets 2 and 4. 
Adapted from Ref.~\cite{Brackley2016NAR}.}
\end{figure}

\begin{figure}
\centerline{\includegraphics[width=15.cm]{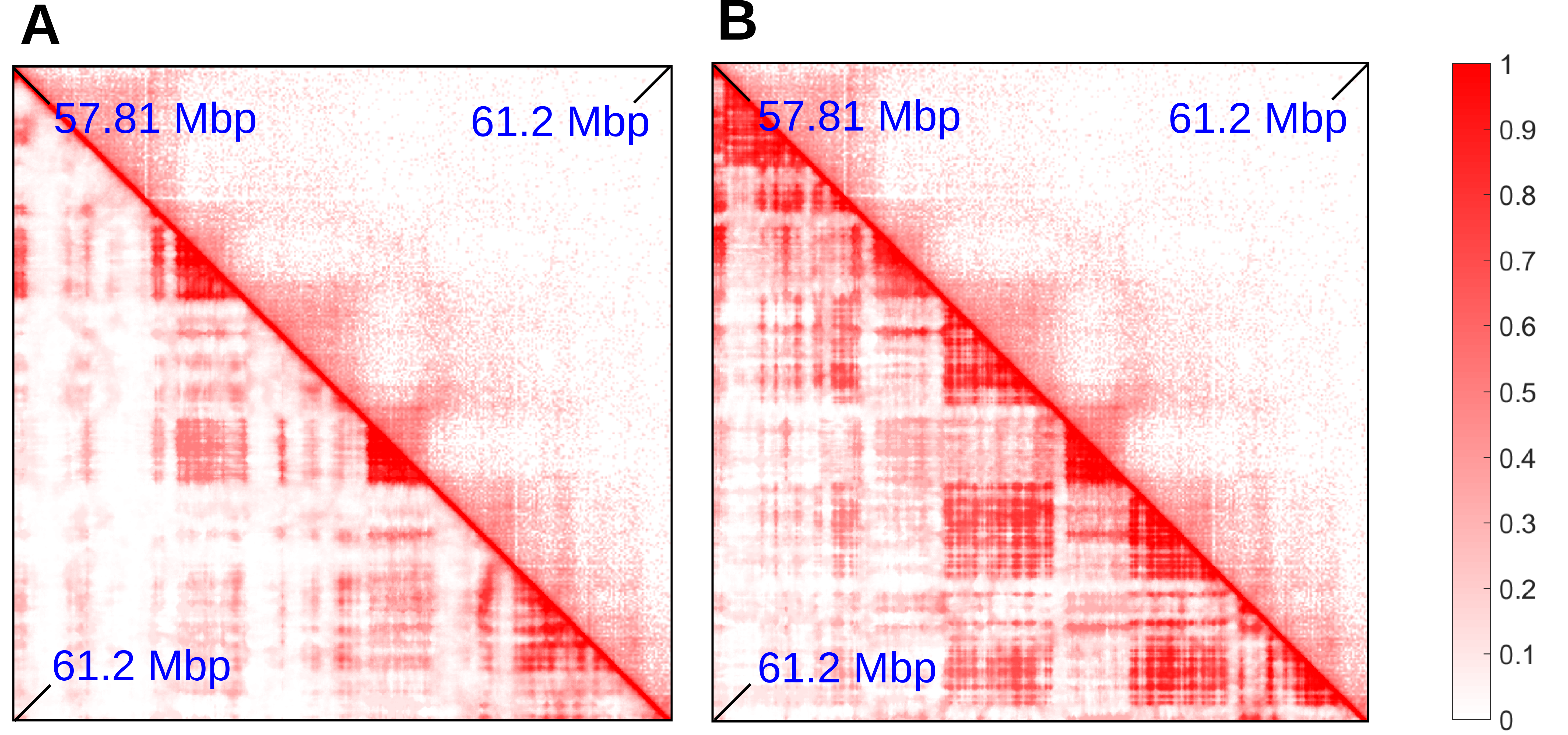}}
\caption{\label{Fig4} 
Adding ``colors'' to the minimal model.
(A,B) Comparison between Hi-C (top triangle) and simulated (bottom triangle) contact maps, for the region between 57.81 and 61.2 Mbp in chromosome 14 in HUVEC cells (coordinates from hg19). Simulations were done similarly to those in Figure 2, and involved 15.5 Mbp of chromatin at 3 kbp resolution, so the region shown is a subset of the whole simulated fragment, chosen to highlight the effect of adding a new species of protein and an additional binding site color to the model. In (A), heterochromatin was colored according to GC content (threshold $\sim$ 40.69\%). It can be seen that several TADs are missing in the simulations. In (B), heterochromatin beads are colored according to H3K9me3 and H3K27me3 tracks (so there are now two possible heterochromatic colors). The latter procedure gives better agreement with the Hi-C data. }
\end{figure}

\end{document}